\documentclass[10pt]{iopart}
\usepackage{iopams}
\usepackage{setstack}
\usepackage[dvips]{graphicx}
\usepackage{epsfig}

\begin{document}

\title{Microscopic energy flows in disordered Ising spin systems}

\author{E. Agliari$^{1,2,3}$, M. Casartelli$^{1,2}$ and A. Vezzani$^{1,4}$}
\address{$^1$ Dipartimento di Fisica, Universit\`a di Parma, Viale
G.P. Usberti n.7/A (Parco Area delle Scienze), 43100 - Parma - ITALY}
\address{$^2$ INFN, gruppo collegato di Parma, Viale G.P. Usberti
n.7/A (Parco Area delle Scienze), 43100 - Parma - ITALY}
\address{$^3$ Albert-Ludwigs-Universit\"at Freiburg, Hermann-Herder-Str. 3, 79104 - Freiburg - GERMANY}
\address{$^4$
S3, National Research Center, CNR-INFM, via Campi 213/a, 41100 - Modena- ITALY}

\begin{abstract}
An efficient microcanonical dynamics has been recently introduced for Ising spin models embedded 
in a generic connected graph even in the presence of disorder i.e. with the spin couplings chosen
from a random distribution. Such a dynamics allows
a coherent definition of local temperatures also when open boundaries are coupled to thermostats, 
imposing an energy flow. Within this framework, here we introduce a consistent definition for local energy currents and we study their dependence on the disorder.
In the linear response regime, when the global gradient between thermostats is small, 
we also define local conductivities following a Fourier dicretized picture. Then, we work out a 
linearized ``mean-field approximation'', where local conductivities are supposed to 
depend on local couplings and temperatures only. We compare the approximated currents 
with the exact results of the nonlinear system, showing the reliability range 
of the mean-field approach, which proves very good at high temperatures and not so efficient in 
the critical region. In the numerical studies we focus on the disordered cylinder but 
our results could be extended to an arbitrary, disordered spin model on a generic 
discrete structures.
\newline
keywords: Transport processes (Theory), Heat conduction, Disordered systems (Theory)
\end{abstract}

%\pacs{05.60-k, 05.50.+q, 44.10.+i}

%\PACS{
%      {05.60-k}{Transport processes}   \and
%      {05.40.Fb}{Microcanonical dynamics} \and
%      {05.40.a}{Fluctuation phenomena, random processes, noise, and Brownian motion}
%      {05.50}{Lattice theory and statistics; Ising problems.}
%     } % end of PACS codes
%end of abstract
%

\section{Introduction}
\label{intro}
The transport properties featured by systems in stationary states, far from equilibrium
is of both theoretical and practical interest: On the one hand, there exist non trivial problems
(e.g. validity bounds of the Fourier description at microscopic scale, 
influence of spatial and topological inhomogeneities featured by the substrate, etc.), which still lack an exhaustive solution; on the other hand, from nanoscales to biological matter, 
the emergence of new materials poses a challenging number of practical problems  
(e.g. fluctuations, clusterizations, correlations, role of  
geometrical irregularities, etc.), where the previous, theoretical approaches 
get an applicative relevance \cite{libro1,libro2,libro3,libro4}.
Discrete models, from simple or interactive random walks to classical spin models, 
play an important role for all such questions \cite{livi,ACV2009,frehland,pippa,acv, acv2,grant,lecomte}. 
In particular, here we deal with an Ising system coupled with thermostats imposing an energy flow, and we study its behavior at microscopic scales. 

The dynamics we adopted in order to simulate the system evolution is a microcanonical dynamics, recently introduced in \cite{ACV2009}, which
has allowed for important enhancements in the study of heat flow in a spin 
lattice coupled to thermostats at arbitrary fixed temperatures. 
The advantages of this dynamics include the simple
definition of a local temperature on each link, the capability of a 
direct  computation of the conductivity (instead of mere diffusivity), the
possibility of applying to an ample variety of discrete spin systems
ranging from models with random couplings to models defined on inhomogeneous 
networks \cite{vivo}.  In particular, the cases of 2-dimensional cylindrical systems, with ordered or
random distributions of couplings $J_{ij}>0$, have  been considered 
in \cite{ACV2009}. The cylindrical geometry of the lattice
suggested to average observables (temperature, 
magnetic energy, etc.) not only in time but also along the ``columns'', 
i.e. periodic rings  at equal distance from the thermostats, and hence orthogonal 
to the flux flowing along ``rows''. This procedure clearly accelerates the numerical 
convergence to well stabilized stationary values. 
In this frame, the validity of the Fourier picture for the energy transport is  
ensured only on the average, by verifying that the average energy
flux through a column is proportional to the average temperature difference
between the sides of the column.
However it is clear that, due to quenched disorder, the local temperature 
fluctuates even within a single column. A natural question is then whether such spatial 
fluctuations influence the local transport properties. 

In this perspective, a basic problem, constituting a non trivial task in itself, 
is a self-consistent definition of microscopic 
currents able to account for the substrate inhomogeneity 
(here generically referring to topological disorder and/or non-constant couplings).
In the present paper we give a consistent solution to this problem which, remarkably, 
applies to arbitrary, connected supports.  
The microscopic link currents originating in each microscopic move can be defined indeed 
independently of the subjacent geometry. 

Afterwards, resuming 
the case of the disordered ferromagnetic cylinder, this very general definition  
is used to check the validity of the Fourier law at microscopic scale. First, 
we verify to this end that, locally, the currents depend on the disorder in a non trivial way. 
More precisely, local currents  and temperatures 
depend not only on the relative coupling $J_{ij}$, but also on the whole configuration of magnetic couplings. Second, we show that, notwithstanding such a complex behavior, a local 
conductivity can be defined on each link. In fact, in a linear response 
regime (i.e. for small enough temperature differences  $\Delta T$ at the borders),
the ratio between the {\sl local}  currents and the {\sl local} temperature gradients 
are independent of $\Delta T$, providing a good definition for the 
local conductivities $K_{ij}$. Clearly, each $K_{ij}$ also depends 
in a non trivial way on the system parameters, its value being indeed
determined  not only by the local coupling and  temperature 
but also by the actual realization of the global disorder. 

Despite such a complex situation, a simple correlation between local conductivities and local couplings 
is present, since on the average the conductivities increase with $J_{ij}$. 
Hence, we  introduced an approximated linear fitting
$K^{\mathrm{MF}}_{ij}=A J_{ij} +B$, where $K^{\mathrm{MF}}_{ij}$ are the approximated 
``mean field'' conductivities and the constants $A$ and $B$ depend 
on the system parameters only, i.e. they are independent of the disorder 
realization.
Within this approach the conduction properties (i.e. the currents) can be 
simply evaluated by solving the discretized linear Fourier equations with 
suitable border conditions.
The approximated currents obtained in this way may be compared with the exact 
results  of the fully non linear system. We evidence that the approach works quite well at 
large enough temperatures, while near $T_c$ (the critical temperature of the equilibrium Ising
lattice) correlations between local conductivities and currents play a much more important role,
 and the local ``mean field'' approach becomes less efficient.
This linear approach for the evaluation of local currents allows for example 
to detect links which, due to the particular disorder realization, are expected to be traversed by large currents and therefore to  
locate the possible critical bonds where the system may fail because of too large a load.

The paper is organized as follows: In Sec.~\ref{microdyn} we briefly review the microcanonical dynamics exploited to simulate the evolution of the system and in Sec.~\ref{sec:currents} we introduce a consistent definition of local currents. Then, in Sec.~\ref{sec:numerics}  we present the numerical results focusing on local currents, on local conductivities and on their dependence on the coupling pattern, while in Sec.~\ref{sec:meanf} we show a mean-field approach to get an estimate of local currents as a function of a given, arbitrary set of couplings; finally Sec.~\ref{sec:concl} is left for conclusions and perspectives. 

\section{Microcanonical Dynamics}
\label{microdyn}
The very novelty of the dynamics introduced in \cite{ACV2009} consists in assigning to the links,
beside the usual magnetic energy $E^m_{ij}$ due to the spin configuration, a ``kinetic'' energy $E^k_{ij}$, 
i.e. a non-negative definite quantity whose variation can compensate the positive or
negative gaps of magnetic energy determined  by the
spin flips of the adjacent nodes. This is clearly different from the Creutz 
microcanonical procedure \cite{creutz1,creutz2}, where the required energy compensation is extracted or
assigned from a bounded amount of energy lying on the nodes themselves: in the latter
case, in fact, both the energy boundness and the connectivity of the nodes could
entail many limitations, ranging from  geometrical or topological constraints 
to the non ergodicity of the system at low energy density.

 Precisely, an elementary move consists in the following:
 \begin{enumerate}
 \item starting from a random distribution of link energies, extract randomly a link
 $(i,j)$ and one of the possible four spin configurations for it;
 
 \item evaluate the variation $\Delta E^m $ of the magnetic energy due to this
choice, checking the whole neighborhood of the link.
Obviously, if couplings are ferromagnetic, being $J_{nk} \in [1- \varepsilon ~,1+ \varepsilon]$ for any link $(n,k)$ and
$\varepsilon > 0$, then $\Delta E^m $ is a real number, and if $\varepsilon = 0$ then $\Delta E^m$ is
an integer whose value depends on the connectivity of the $k$ and $n$ nodes;
 \item if $\Delta E^m \leq 0$, accept the choice and increase the link kinetic energy $E^k_{ij}$ of $\Delta E^m $; 
\item if $\Delta E^m > 0$, accept the choice and decrease the link kinetic energy of $\Delta
E^m $ only if the link energy remains non negative.
  \end{enumerate}

The unit time step will be a series of $N$ moves, where $N$ is the number of links.
Since a link is defined only by the adjacent nodes, and nothing in the rule above
refers to a definite structure (e.g. a lattice), the move is implementable on every non-oriented
connected graph. This ensures the great generality of the dynamics.

Starting from the observation that magnetic and kinetic energies behave as non
correlated observables, in \cite{ACV2009} many points have been  supported by theoretical arguments
and tested numerically, for both homogeneous and disordered
links. In particular, at equilibrium the Boltzmann distribution is recovered
and the system is ergodic at all temperatures.
This opens the possibility to extend the definition of temperature $T_{ij} $ as a {\sl local}
observable, in fact a {\sl link} observable, which recovers the averaged  kinetic energy
$ \langle E_{ij}^k \rangle$. This holds also in non-equilibrium, stationary
states, i.e. states forced by thermostats at different temperature.

As for thermostats, their definition requires in general the presence of contact
borders, which for the cylinder are the first and last columns. In this case, every thermostat  consists in a number of additional columns (2 are
enough) regulated at every step by the usual Metropolis equilibrium dynamics with
the wanted inverse temperature $\beta$.

\section{Definition of Local Currents}
\label{sec:currents}

\begin{figure}[tb] \begin{center}
\includegraphics[width=.90\textwidth]{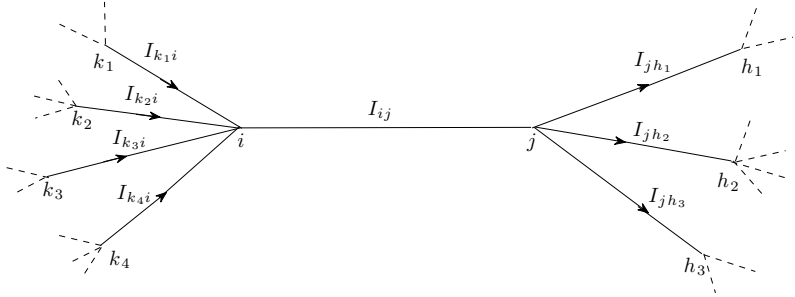}
\caption{\label{fig:correnti} Schematic representation of currents for an arbitrary link belonging to an arbitrary graph. In this example the link considered is the one labelled as $(i,j)$ with $z_i=4$ while $z_j=3$. The arrows indicates the direction attributed to the energy flow on each link.}
\end{center}
\end{figure}

In order to calculate local conductivities also in the presence of disorder (due to either topological and coupling inhomogeneity) it is necessary to define a current for each link. We now introduce the scheme through which we are able to consistently assign a current to any arbitrary link in the structure considered. Such a scheme can be applied to a generic structure, as envisaged in Fig.~\ref{fig:correnti}.

In our dynamics energies are naturally assigned to each links; however in order
to define link currents it is useful to assign one half of the link energy
to each of its relevant nodes. Let us consider the link $(i,j)$ connecting $i$ and $j$; Let $V_i$ and $V_j$
their neighborhoods,  $z_i = \mid V_i \mid $ and $z_j = \mid V_j\mid $ their respective 
coordinations. Then,  $k \in V_i$ and $h \in V_j$ are 
the neighbouring site labels for $i$ and $j$, respectively. 
We fix a direction for currents on adjacent links: currents on links connected 
through $i$ are incoming while those connected through $j$ are outgoing.

Now, given a spin flip involving the link under consideration, namely either 
the $i$-th spin, or the $j$-th spin or both, for adjacent links as well as 
for $(i,j)$ we have a possible energy variation denoted with 
$\Delta E_{i k}$, $\Delta E_{j h}$ and $\Delta E_{i j}$
respectively. For links $(i,k)$ and $(h,j)$ half of such 
variations contributes to the current on the link itself since it 
represents the energy flow outgoing and incoming from the sites $k$ and 
$h$. Therefore we have
$$
I_{ki} = - \frac{\Delta E_{ki}}{2}, ~~~
I_{jh} = + \frac{\Delta E_{jh}}{2}, 
$$ 
where different signs derives from the flow direction we have chosen (see Fig.~\ref{fig:correnti}).

As for $I_{ij}$ and the energy variation $\Delta E_{ij}$, they satisfy the following
\begin{eqnarray}
\sum_{k\in V_i} I_{ki} - I_{ij} = \sum_{k\in V_i} \frac{\Delta E_{ki}}{2} + \frac{\Delta E_{ij}}{2}, 	\\
- \sum_{h\in V_j} I_{jh} + I_{ij} = \sum_{h\in V_j} \frac{\Delta E_{jh}}{2} + \frac{\Delta E_{ij}}{2}.
\end{eqnarray}
 
In the above equations, the left-hand side represents the currents 
arriving and departing from $i$ (and $j$) while the right-hand side is 
the consequent energy variation.
Hence, we get $\Delta E_{ij} = - \sum_{k\in V_i} \Delta E_{ki} - \sum_{h\in V_j} \Delta E_{hj}$, as consistent with energy conservation, and
\begin{equation}
I_{ij} = \frac{1}{2} \left [  - \sum_{k\in V_i} \Delta E_{ki} + \sum_{h\in V_j} \Delta E_{jh} \right ].
\end{equation}

This scheme works for any arbitrary topology and, of course, even in the presence of a disordered distribution of couplings: It only requires the knowledge of the local energy variations consequent to any spin-flip.

\section{Numerical results}\label{sec:numerics}
In the following we report and discuss the results obtained by means of Monte Carlo simulations performed on squared cylinders
of $N= L\times L$ sites,  for different realizations of disorder (encoded by the $N\times N $ matrix $\mathbf{J}$). The first and last columns of $L$ sites are open, in contact with thermostats at temperatures $T_1$ and $T_2$ (to fix ideas $T_1 < T_2$), and different choices of temperatures are considered. Since we are interested in local quantities, we especially focus on small sizes, which allow fast thermalization though displaying the relevant features of the non-equilibrium behavior (see \cite{ACV2009}).  

First of all, let us consider local currents and local temperatures.
For a single realization of the disorder in the window $(1-\epsilon , 1+\epsilon )$, currents $I_{ij}$ may be calculated according to the scheme described in the previous section; one can also measure the local temperatures $T_{ij}$ which, for randomly distributed couplings equal the relevant average kinetic energy $\langle E_{ij}^k \rangle$ \cite{ACV2009}.
Then, from such local temperatures $T_{ij}$, we can estimate a temperature $T_i$ to associate to each node, namely
\begin{equation}
T_{i} = \frac{1}{|V_i|} \sum_{j \in V_i} T_{ij},
\end{equation}
Therefore, the local temperature gradient among nodes $i$ and $j$ is naturally given by $\Delta T_{ij} = T_{i} - T_{j}$.
  
\begin{figure}[tb] \begin{center}
\includegraphics[width=.80\textwidth]{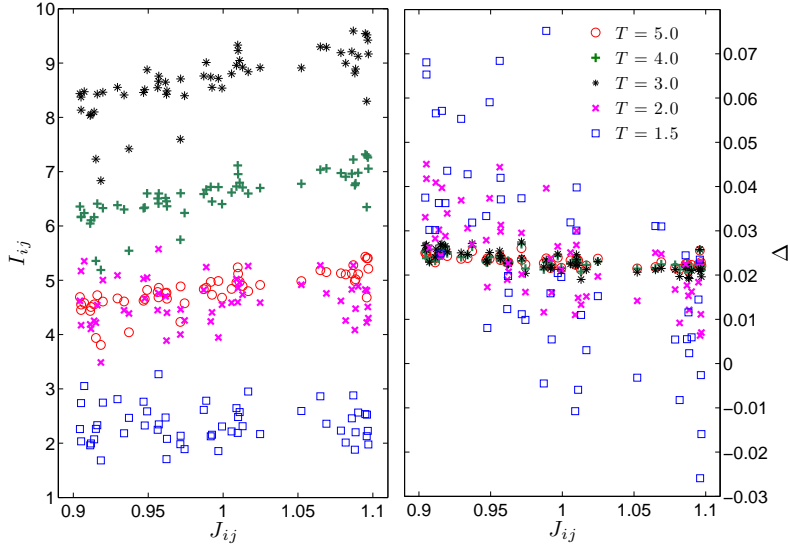}
\caption{\label{fig:currT} Local currents $I_{ij}$ (left panel) and local temperature gradients $\Delta T_{ij}$ (right panel) versus the corresponding coupling $J_{ij}$ for a cylindrical lattice with linear size $L=10$, $\Delta T =0.2$ and different choices for $T_2$, as shown by the legend. The coupling pattern $\mathbf{J}$ is the same for all sets of data points depicted.} 
\end{center}
\end{figure}

Numerical data for the average currents $\langle I_{ij} \rangle$ and $\Delta T_{ij}$, as a function of the pertaining coupling strength $J_{ij}$, are shown in Fig.~\ref{fig:currT}; similar results are obtained for different realizations $\mathbf{J}$. We notice that, being $\Delta T = T_2 - T_1$ the global difference of temperature, local gradients are distributed around the expected value $\Delta T / L$, with large spread especially for low temperatures $T_1, T_2$ and a slight correlation with the pertaining couplings, that is, large interaction strengths $J_{ij}$ correspond to smaller gradients $\Delta T_{ij}$. Local currents display a larger degree of correlation: large interaction strength $J_{ij}$ correspond to larger magnitudes for currents $\langle I_{ij} \rangle$. We also notice that different temperatures for thermostats give rise to similar, though shifted, distributions of data points. Other realizations of the same disorder give, of course, different point distributions, but the linear interpolation and the value of the  fluctuations prove to be very robust, so that the definition of {\sl interpolating} currents $\bar{I}_{ij}$ is reliable at every fixed $\epsilon$ and $T_1, T_2$.
It is also noteworthy that, for a fixed $\Delta T$, currents are not monotonic in $T$: referring to Fig.~\ref{fig:currT} (left panel), their magnitude is maximum at $T=3$, i.e. around the critical temperature expected for the (disordered) two-dimensional Ising model \cite{ACV2009}. 

\begin{figure}[tb] \begin{center}
\includegraphics[width=.80\textwidth]{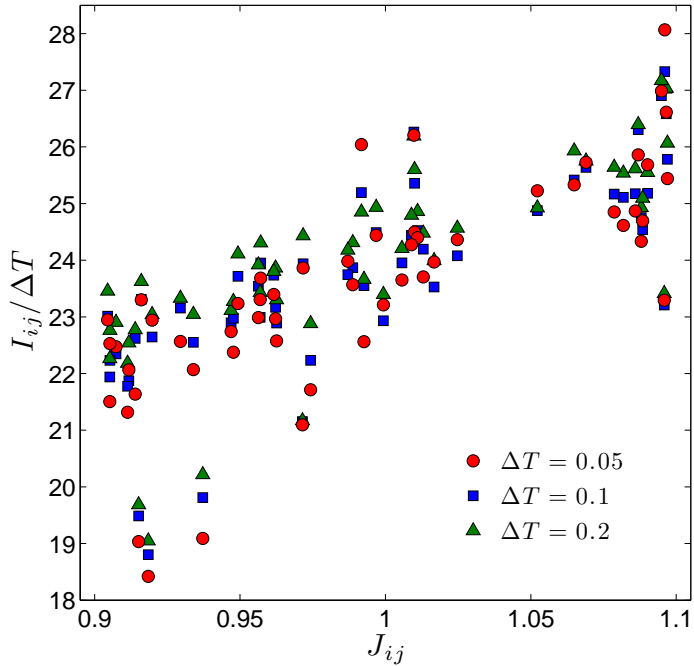}
\caption{\label{fig:LinResp} Local currents $I_{ij}$ divided by the global temperature difference $\Delta T$ as a function of the pertaining coupling $J_{ij}$. As explained by the legend different gradients, namely $\Delta T = 0.05$, $\Delta T = 0.1$ and $\Delta T =0.2$ are compared, while $T_2=5$, the coupling pattern $\mathbf{J}$ and $L=10$ are kept fixed. Notice that the average difference between data pertaining to $\Delta T = 0.05$ and $\Delta T = 0.1$, and data pertaining to $\Delta T = 0.1$, $\Delta T = 0.2$ is, respectively, $0.3$ and $0.4$, therefore below the error $\approx 4 \%$.}
\end{center}
\end{figure}

Moreover, when $\Delta T \ll 1$, the linear response theory holds: Both the local currents $\langle I_{ij} \rangle$ and the local temperature gradients $\Delta T_{ij}$ are proportional to the global difference of temperature $\Delta T$, as corroborated by the collapse of data points in Fig.~\ref{fig:LinResp}, where the values relevant to different temperatures are compatible with the numerical error. 
In this perspective, given local currents and local gradients, we can introduce the local conductivities according to
\begin{equation}
\label{cond}
K_{ij}= \frac{\langle I_{ij} \rangle} {T_i-T_j},
 \end{equation}
which are well defined quantities describing the microscopic conduction of the 
system. We expect that a similar definition works as well for generic topologies, at least in the regime of small gradients.

From energy conservation and equation (\ref{cond}) we obtain the local 
Fourier equation characterized by link dependent conductivities 
\begin{equation}
\label{fou}
\sum_{j \in V_i} K_{ij} (T_i-T_j)= -\frac{\partial \langle E_i \rangle}{\partial t},
 \end{equation}
where $\langle E_i \rangle=1/2 \sum_{j \in V_i} \langle E_{ij} \rangle$ is the total energy relevant to site $i$; the continuum notation is used for convenience, with the usual warning about the meaning of derivatives in these discrete-time systems (see for instance \cite{cmv}). In general terms, the expression in Eq.~\ref{fou} describes a system where an external field, or gradient, along one axis and a fluctuating local field, or disorder, have been applied; the former makes the temperature increase by a constant amount per row of nodes, while the latter gives rise to currents non-trivially depending on the whole environment. Indeed, the same equation is also used in the context of random resistor networks \cite{kirk}, where the voltage and the conductance play the role of the temperature and of the conductivity, respectively.

We remark that  conductivities $K_{ij}$ depend on system parameters in a very complex way, indeed their values is determined by the temperature $T$, by the degree of disorder $\epsilon$ and by the whole coupling pattern $\mathbf{J}$. Moreover, we verified that the dependences on the three arguments are intrinsically interplaying, namely that given $K_{ij} = f(\mathbf{J}, T, \epsilon)$, factorizations like $f(\mathbf{J}, T, \epsilon) = f_1(\mathbf{J}) \cdot f_2(T,\epsilon)$ are ruled out.

\begin{figure}[tb] \begin{center}
\includegraphics[width=.80\textwidth]{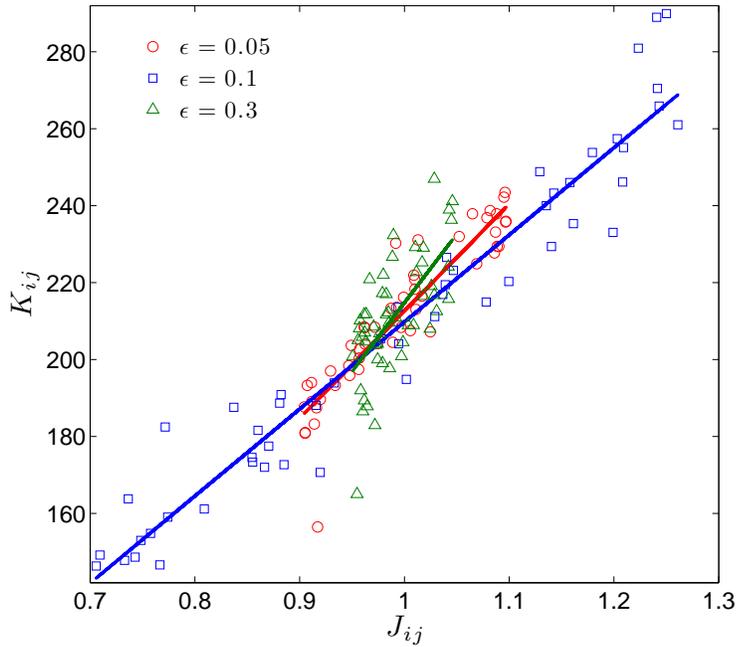}
\caption{\label{fig:Eps} Local conductivities $K_{ij}$ versus pertaining local couplings $J_{ij}$ for a square cylinder with linear size $L=10$ and thermostats at temperatures $T_1=4.8$ and $T_2=5.0$, respectively; different degrees of disorder $\epsilon = 0.05$, $\epsilon = 0.1$ and $\epsilon = 0.2$, have been considered and represented with different symbols, as shown in legend. Linear curves represent the best fits, whose angular coefficients are approximately $358$, $279$ and $226$ respectively.}
\end{center}
\end{figure}

As shown in Fig.~\ref{fig:Eps}, local conductivities are correlated on the average with the local couplings, and such a correlation gets stronger (the fitting curve has larger slope) for smaller values of $\epsilon$.
The ``ordered system limit'', i.e. $\epsilon \to 0 $, is in a sense singular, since the distributions of the $J_{ij}$ shrinks in a single point. 
  
\begin{figure}[tb] \begin{center}
\includegraphics[width=.80\textwidth]{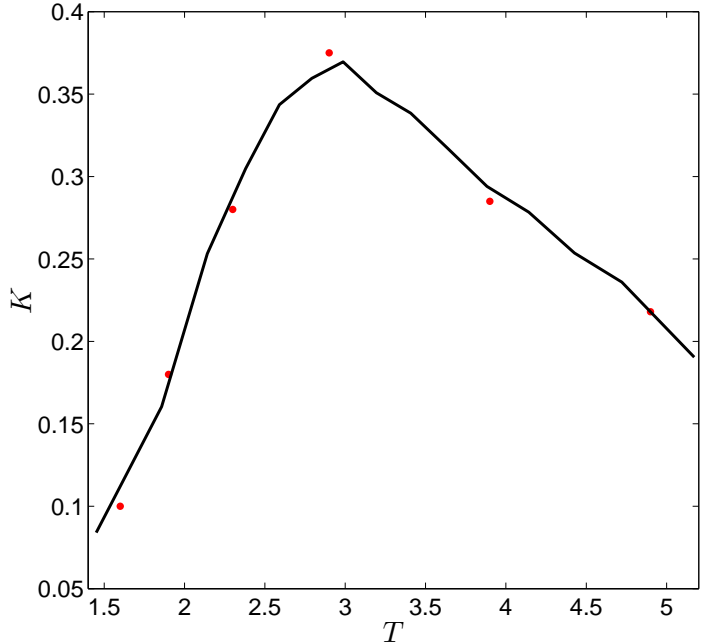}
\caption{\label{fig:Confronto} Comparison between average local conductivities measured according to Eq.~\ref{eq:average} ($\bullet$) and according to a ``mesoscopic'' measure of conductivity (continuous line) \cite{ACV2009}, respectively, as a function of temperature.}
\end{center}
\end{figure}

Now, from such local conductivities it is possible to derive an estimate for the conductivity $K(T, \mathbf{J})$ expected for a system at a temperature $T$ and in the presence of disorder $\mathbf{J}$, by assuming $\Delta T \ll 1$ and averaging over all local conductivities:
\begin{equation} \label{eq:average}
K(T,\mathbf{J}) = \frac{1}{N} \sum_{i} \sum_{j \in V_i} K_{ij}. 
\end{equation}
%where $N$ is the total number of links.
In Fig.~\ref{fig:Confronto} we compare such measures realized at different temperatures with a ``mesoscopic'' measure of $K$ based on the heat flow passing from one layer to the next one in a similar cylinder \cite{ACV2009}.
The very good agreement between the two estimates provides a further confirmation about the consistency of our definition of local currents and conductivities.

\section{Mean Field Approach}
\label{sec:meanf}
Within the linear response approach, local temperatures can be evaluated by solving the Fourier equation (\ref{fou}) 
imposing the stationarity of local energy at every node:
\begin{equation} \label{risplin}
\sum_{j \in V_i} K_{ij}(T_i - T_j)=0.
\end{equation}
Suitable boundary conditions should be chosen forcing the temperatures at the borders to be fixed at $T$ and $T+\Delta T$ respectively. Then currents can be evaluated from temperatures as $\langle I_{ij} \rangle =K_{ij}(T_i-T_j)$. Clearly, Eq.~\ref{risplin} is useless for practical calculations, as $K_{ij}$ has to be obtained from the (numerical) solution of the whole spin dynamics. Moreover, we have already evidenced that $K_{ij}$'s depend on the whole configuration of local couplings $J_{ij}$ in a  non trivial way. However Fig.~\ref{fig:Eps} suggests that a simple ``mean field'' approximation should be possible imposing $K_{ij}$ to be dependent only on the local coupling $J_{ij}$, i.e. the mean-field, local conductivities are defined as $K^{\mathrm{MF}}_{ij}=A J_{ij} + B$, where  $A$ and $B$ are the fitting parameters used in Fig.~\ref{fig:Eps} depending on $T$ and $\epsilon$ only. Clearly, this approach is much simpler since, once $A$ and $B$ are known, the mean-field, local conductivities can be inferred for any realization of the disorder. Then, using the linear equations (\ref{risplin}), one can obtain mean-field, local temperatures $\sum_{j \in V_i} K^{MF}_{ij}(T^{MF}_i-T^{\mathrm{MF}}_j)=0$ and currents $I^{MF}_{ij}=K^{\mathrm{MF}}_{ij}(T^{MF}_i - T^{MF}_j)$. 

\begin{figure}[tb] \begin{center}
\includegraphics[width=.80\textwidth]{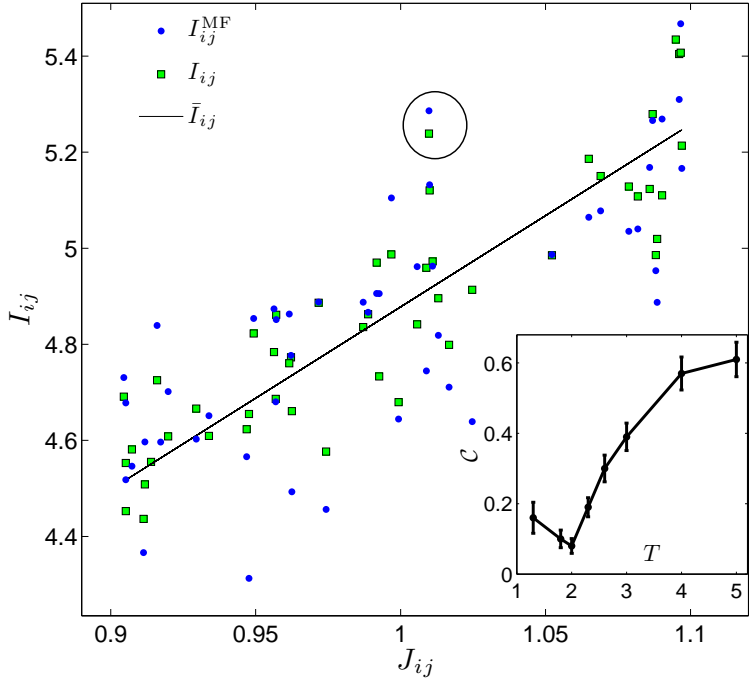}
\caption{\label{fig:corr} Main figure: $\bar{I}$ (line), $I^{MF}$ ($\square$) and $I_{nk}$ ($\bullet$) versus the related coupling strength for $T_1 = 4.8$ and $T_2=5$; analogous results are found also for other temperatures. Inset: temperature dependence for the correlation parameter $\mathcal{C}$.} 
\end{center}
\end{figure}

In Figure \ref{fig:corr} we evidence that local, mean-field currents are indeed a good approximation of the exact results. More precisely, once the local deviations $\phi_{ij}$ and $\phi_{ij}^{\mathrm{MF}}$ are defined as
\begin{eqnarray}
\phi_{ij} = \langle I_{ij} \rangle - \bar{I}_{ij},\\
\phi_{ij}^{\mathrm{MF}} = I_{ij}^{\mathrm{MF}} - \bar{I}_{ij},
\end{eqnarray}
we can quantify the correlation between the real values $\langle I_{ij} \rangle$ of currents, i.e. those obtained from numerical simulations, and the estimate values 
$I_{ij}^{\mathrm{MF}}$, i.e. those obtained from the mean-filed approach, by means of the correlation coefficient
\begin{equation}
\mathcal{C}=
\frac{\mathbb{E} ( \phi_{ij} \cdot \phi_{ij}^{\mathrm{MF}} ) -\mathbb{E} \phi_{ij} \mathbb{E} \phi_{ij}^{\mathrm{MF}} }
{\sqrt{ \left[ \mathbb{E} (\phi_{hn}^{\mathrm{MF}})^2  -
(\mathbb{E} \phi_{ij}^{\mathrm{MF}})^2 \right] }},
\end{equation}
where averages $\mathbb{E}$ are obtained summing over the whole set of links and  $\bar{I}_{ij}$ is the value of the local current obtained with the linear fit of Figure \ref{fig:currT}. Notice that $\mathcal{C}$ ranges from $-1$ (anticorrelation) to $+1$ (correlation) and $\mathcal{C}=0$ means no correlation. We measured the quantity $\mathcal{C}$ finding strictly positive values for all the temperatures considered. The positivity of correlation evidences that $I^{\mathrm{MF}}_{ij}$ approximates currents $\langle I_{ij} \rangle$ better than $\bar{I}_{ij}$. This remarkable property is easily explained since the mean field field approach not only takes into accounts  the correlations between local couplings, currents and conductivities evidenced in Figs.~\ref{fig:currT}-\ref{fig:Eps}, but also takes into account the local conservation of energies encoded in Eq.~\ref{risplin} and representing one of the basic features of the microscopic spin dynamics. In particular, on the contrary of $\bar{I}_{ij}$, currents $I^{MF}_{ij}$ are conserved at every node.
It is also worth noting that larger values of $\mathcal{C}$, and therefore a better efficiency of the the mean-field approximation, are found for large temperatures. For instance, at $T=5$ we get $\mathcal{C} = 0.63$; as the temperature is lowered the correlation decreases displaying a possible minimum around $T_c$. Critical effects apart, the mean-field approach seems to provide good estimates especially for large temperatures: indeed for $T$ approximately larger than $3.5$ one has $\mathcal{C} > 0.5$ and we checked that  at $T=500$, $\mathcal{C}$ is close to $0.94$. We also underline that $\mathcal{C}=1$ means that local conductivities are purely local quantities, independent of the neighborhood.

We remark that the links crossed by large currents may be sharply identified within this approximation, which, for example, captures  the link evidenced in Fig.~\ref{fig:corr}: although its intermediate coupling value, it carries a large current. In other terms, our approximation is able to locate the regions characterized by high currents which, in realistic realizations, could lead to a failure of the link itself.

\section{Conclusions and Perspectives}
\label{sec:concl}

In this work we addressed the general problem of a spin model on arbitrary discrete structures with a microcanonical dynamics, focusing on the conduction properties at microscopic scales. First, we introduced a consistent definition of local currents. Then we applied such general definition in the case of  disordered ferromagnetic Ising model on a cylindrical structure, where  boundaries are coupled with thermostats at different fixed temperatures.
We highlighted that the local microscopic currents depend non trivially on the whole distribution of quenched couplings and  a consistent definition of local conductivity has been introduced, at least in regime of linear response, i.e. when the temperature gradient $\Delta T$ is small.

In spite of the aforementioned dependence of local quantities (currents, temperatures and conductivities) on the whole disorder arrangement, numerical results evidenced a simple correlation between local conductivities and the pertaining local couplings. These correlations suggested the development of  an approximated mean-field like approach, where local conductivities depend on local couplings only, and the conduction properties (i.e. the currents) can be easily  evaluated by solving the discretized linear Fourier equations with suitable border conditions. Such approximation is especially effective at large temperatures. With respect to previous work on the cylindrical spin model, the strict requirement on the smallness of $\Delta T$ is due precisely to the fact that in the present case we focus on the microscopic aspects. 

This pattern of results naturally indicates preferential lines of future developments. 
First of all, the definition of currents holds on a general geometrical substrate. In   
a complex topological structure, some aspects of the framework we have worked out  are expected 
to be still valid, in particular the presence of a linear response 
regime and the definition of local conductivities whose values may depend,
however, in a non trivial way on the topology of the underlying structure.

Moreover, an important point to investigate concerns the extent of correlation length, as a function 
of temperature, degree of disorder and topology. In other terms, it would be interesting to determine 
the length of the radius such that the external pattern constitutes a practically uniform background, 
without any influence on the local currents and conductivities. Also the response of local conductivity 
to small coupling perturbations would be in order. This kind of problems, at finite and low temperatures, 
has obviously to do with the correct definition of the mesoscopic scale we spoke about. At $T\gg 1$, 
analytical and numerical estimates say that this radius possibly reduces to one link, which means that the conductivity is a local property.

We also infer the interest of a more precise characterization of the behaviour of the system both at low 
temperatures and for  $T \approx T_c$, where the linear approximation has some difficulties, possibly not 
only of numerical nature. Finally, the whole set of results clearly indicates that we have an extremely 
flexible calculation engine ready for more complex realizations of disorder, including the case of null 
and negative couplings, i.e. diluted and spin-glass systems.

\end{document}